\documentclass{article}

\usepackage{arxiv}

\usepackage[utf8]{inputenc} 
\usepackage[T1]{fontenc}    
\usepackage{url}            
\usepackage{booktabs}       
\usepackage{amsfonts}       
\usepackage{nicefrac}       
\usepackage{microtype}      
\usepackage{lipsum}		
\usepackage{graphicx}
\usepackage{natbib}
\usepackage{doi}
\usepackage{lineno}
\usepackage{amsmath}
\usepackage{hyperref}       

\modulolinenumbers[5]

\title{Self-citation Analysis using Sentence Embeddings}

\author{ \href{https://orcid.org/0000-0002-1979-3915}{\includegraphics[scale=0.06]{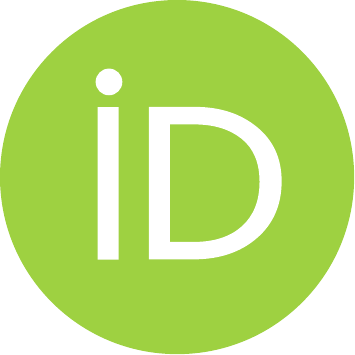}\hspace{1mm}Athanasios Lagopoulos}\\
	School of Informatics\\
	Aristotle University of Thessaloniki\\
	Thessaloniki, Greece \\
	\texttt{lathanag@csd.auth.gr} \\
	\And
	\href{https://orcid.org/0000-0002-7879-669X}{\includegraphics[scale=0.06]{orcid.pdf}\hspace{1mm}Grigorios Tsoumakas} \\
	School of Informatics\\
	Aristotle University of Thessaloniki\\
	Thessaloniki, Greece \\
	\texttt{greg@csd.auth.gr} \\
}

\date{}

\renewcommand{\headeright}{}
\renewcommand{\undertitle}{}
\renewcommand{\shorttitle}{Self-citation Analysis using Sentence Embeddings}

\hypersetup{
pdftitle={Self-citation Analysis using Sentence Embeddings},
pdfsubject={q-bio.NC, q-bio.QM},
pdfauthor={Athanasios Lagopoulos,Grigorios Tsoumakas},
pdfkeywords={Self-citation, Semantic similarity, Sentence embedding},
}

\begin{document}
\maketitle

\begin{abstract}
	The purpose of citation indexes and metrics is intended to be a measure for scientific innovation and quality for researchers, journals, and institutions. However, those metrics are often prone to abuse and manipulation by excessive and unethical self-citations induced by authors, reviewers, editors, or journals. Identifying whether there are or not legitimate reasons for self-citations is normally determined during the review process, where the participating parts may have intrinsic incentives, rendering the legitimacy of self-citations, after publication, questionable. In this paper, we conduct a large-scale analysis of journal self-citations while taking into consideration the similarity between a publication and its references. Specifically, we look into PubMed Central articles published since 1990 and compute similarities of article-reference pairs using sentence embeddings.  We examine journal self-citations with an aim to distinguish between justifiable and unethical self-citations.
\end{abstract}

\keywords{Self-citation \and Semantic similarity \and Sentence embedding}

\section{Introduction}
Different forms of self-citation can be observed in scientific publishing. Direct self-citation, where authors cite their previous work, are the ones most associated with the term. Co-author and collaborative self-citation come as an extension to direct self-citation adding more complexity to the self-citation network. Coercive citation, where reviewers ask authors to include references to their work, are considered the most difficult to detect, and along with the journal/institution self-citations they complete the list of the most common types of self-citation \citep{Ioannidis2015}.

The scientific impact of an individual, an institution, or a publisher is typically determined by quantitative metrics and indices \citep{Wren2020}. Citation count is the prevalent metric of recognition in science, while qualitative metrics are discarded as hard to formulate and employ \citep{fortunato2018science}. Promotions and awards are granted to highly cited researchers, and journals with high impact factors are considered to be the most appealing and prestigious ones. Yet, most of the metrics such as citation count, h-index, and impact factor do not investigate self-citations. Thus, efforts to game or hack such metrics are made to increase the visibility of a research piece and consequently to act as advertisement and recognition for the author/publisher/institution \citep{Szomszor2020, Copiello2019}.

Self-citation, and especially excessive self-citation, are usually conceived by researchers as an unethical practice. However, recklessly criticizing self-citations is unfitting to the research community. Authors usually utilize their prior work to develop novel approaches and a high self-citation ratio (i.e. the number of self-citations to the total number of citations) may signify their part in science. Meanwhile, in narrow field journals, self-citations are fundamental and sometimes necessary \citep{Szomszor2020}. The legitimacy of self-citation should be questioned when patterns are identified and dubiously relevant or completely irrelevant references are included \citep{Humphrey2019, Wren2019}.

Peer review is the process responsible for checking the validity of self-citations, but it has been judged for bias, inconsistency, corruption, and reluctance on raising concerns on single incidents \citep{Thombs2015, Wren2020,martin2013whither}. Different suggestions for transparency have been made, such as stricter review rules and open peer reviews, but pros and cons are still debated. From the scope of metrics, austere measures are employed with the complete removal of self-citations which can be useful only for extreme cases \citep{Ioannidis2015}.

This paper introduces a novel way for detecting potentially unethical self-citations based on the semantic similarity of a publication and its references. The underlying assumption is that articles mostly reference other articles that research related subjects and belong to the same field of study. State-of-the-art sentence embeddings are adopted to estimate the similarity of article-reference pairs. A large-scale analysis of journal self-citations on articles indexed by PubMed shows promising results on detecting unethical practices incited by journals. Our method does not blindly punish self-citations. Moreover, we introduce the ReLy (Relevance Legitimacy) score that influences the self-citation ratio of a publication by considering the semantic similarity of publication-reference pairs. A list of more than 3800 journals with their ReLy score is made publicly available for further analysis.

The rest of the paper is structured as follows. After  a  discussion  of  the related work and sentence embeddings in Section \ref{sec:related}, we present our methodology in Section \ref{sec:methodology}. In Section \ref{sec:results}, we demonstrate and discuss the results of our analysis. Finally, we identify the limitations of this study and conclude by drawing future directions in Sections \ref{sec:limitations} and \ref{sec:conclusion}, respectively. 

\section{Related Work}
\label{sec:related}
We first provide related studies that analyze self-citations in academic publishing and then a brief introduction to word and sentence embeddings that we use in our analysis.

\subsection{Self-citation Analysis}

Several research papers and reports have been published in the past analyzing self-citation networks and their impact on scientometrics and research visibility. Early studies were focused on author self-citations and the motivation behind them \citep{bonzi1991motivations}, their regularities \citep{aksnes2003macro,Glanzel200463}, their influence on academic indicators \citep{fowler2007does,glanzel2004influence,van2008self}, and their influence on credibility and recognition in academic publishing \citep{hyland2003self}. Moreover, several studies extended those concerns to co-author self-citations \citep{glanzel2004does,schubert2006weight,costas2010self,lin2012relationship,peroni2020practice}.

Most of the above studies were concerned with defining self-citations and the incentive behind them. Recent studies focus more on detecting self-citations \citep{yang2016detecting} and their manipulation techniques, not only by authors \citep{fong2017authorship}, but also by journals \citep{heneberg2016excessive} or even countries \citep{di2019open,baccini2019citation,seeber2019self,peroni2020practice}. To that extend, some studies have performed simulations \citep{bartneck2011detecting,lopez2012manipulating}, many proposed new metrics, indices \citep{flatt2017improving,majeed2019self,ioannidis2019standardized,ioannidis2019user,kacem2020tracking,simoes2020self} or graphical procedures \citep{Szomszor2020}, and others looked into the direct impact on publications' visibility \citep{gonzalez2019journal,Copiello2019}. 

Furthermore, several studies have centered their attention on journal self-citation.  \citet{gazni2020journal} analyzed the journal self-citation trends from 1975 to 2017 and contrasted them with the usefulness of self-citation indicators. Other studies examined the issue of coercive citations \citep{Wren2019, Humphrey2019}, which are the hardest to detect. Their focus was on gathering opinions on that matter \citep{wilhite2012coercive}, manually looking at the extend of the issue \citep{Thombs2015} and automatically detecting journals with classification methods \citep{yu2014classification}. 

Finally, few studies have looked into the context surrounding self-citations. \citet{zhou2020self} probed into the motives and context of citations by examining them via the lenses of polarity, density, and location in the text. \citet{galvez2017assessing} empirically studied the role of author self-citations as a mechanism of knowledge diffusion. He compared the semantic content shared between the article and the (non-)self-citations using Latent Dirichlet Allocation (LDA). He concludes that authors do not cite themselves in an irrelevant way and they usually cite articles close to their content. To our knowledge, this is the only study that makes use of semantic information to analyze self-citations. 

\subsection{Word and Sentence Embeddings}
In the past decade, word embeddings have gained great attention within the Natural Language Processing (NLP) area due to their high performance in different tasks. The term word embeddings was first introduced by \citet{bengio2003neural}, but it was much later that they were established as a powerful tool \citep{collobert2008unified}. Finally, it was the introduction of word2vec \citep{mikolov2013efficient} and the rapid expansion in computational power that brought word embeddings into the mainstream.

Word embeddings are based on the idea that contextual information alone constitutes an important representation of linguistic terms. Words that occur in similar contexts tend to have similar meanings. This link between similarity in how words are distributed and what they mean is called the distributional  hypothesis. In the most basic form, their architecture constitutes of a logistic regression model that distinguishes between neighboring context words (positive examples) and randomly sampled words in the lexicon (negative examples). The learned weights of the model are then used as the embeddings. Thus, words are represented as semantically rich $n$-dimensional vectors which can then be used in a vast range of NLP applications. A detailed description of such architecture is out of the scope of this study. 

The success of word2vec urged the scientific community to expand these architectures remarkably achieving state-of-the-art results in many NLP tasks such as word similarity, word analogy, text classification, and others. Some of the most used architectures are Glove~\citep{pennington2014glove}, fastText~\citep{bojanowski2017enriching} and architectures that use transformers, such as BERT~\citep{devlin2018bert} and GPT~\citep{radford2018improving}, which achieve impressive results in question answering and machine translation, among other tasks. 

In the meantime, several sentence embedding techniques were developed to eliminate the need for semantically meaningful representations for not just words but also phrases and documents. Most of the proposed approaches, combine word embeddings or expand them to sentence embeddings \citep{le2014distributed,kusner2015word,arora2017asimple,conneau2017supervised,pagliardini2017unsupervised,reimers2019sentence}. Sentence embeddings achieve state-of-the-art results in semantic textual similarity tasks, natural language inference, and even in movie and product review classification tasks. Our work explores the use of sentence embeddings to capture the semantic similarity of articles found in the biomedical literature with the aim to analyze journal self-citations.

\section{Methodology}
\label{sec:methodology}

This section presents our methodology for analyzing journal self-citations based on semantic similarity. First, we present the dataset we use. Then, we give details on journal self-citation in our dataset and finally we introduce the ReLy (Relevance Legitimacy) score. 

\begin{figure}[!h]
    \centering
    \includegraphics[width=0.55\textwidth]{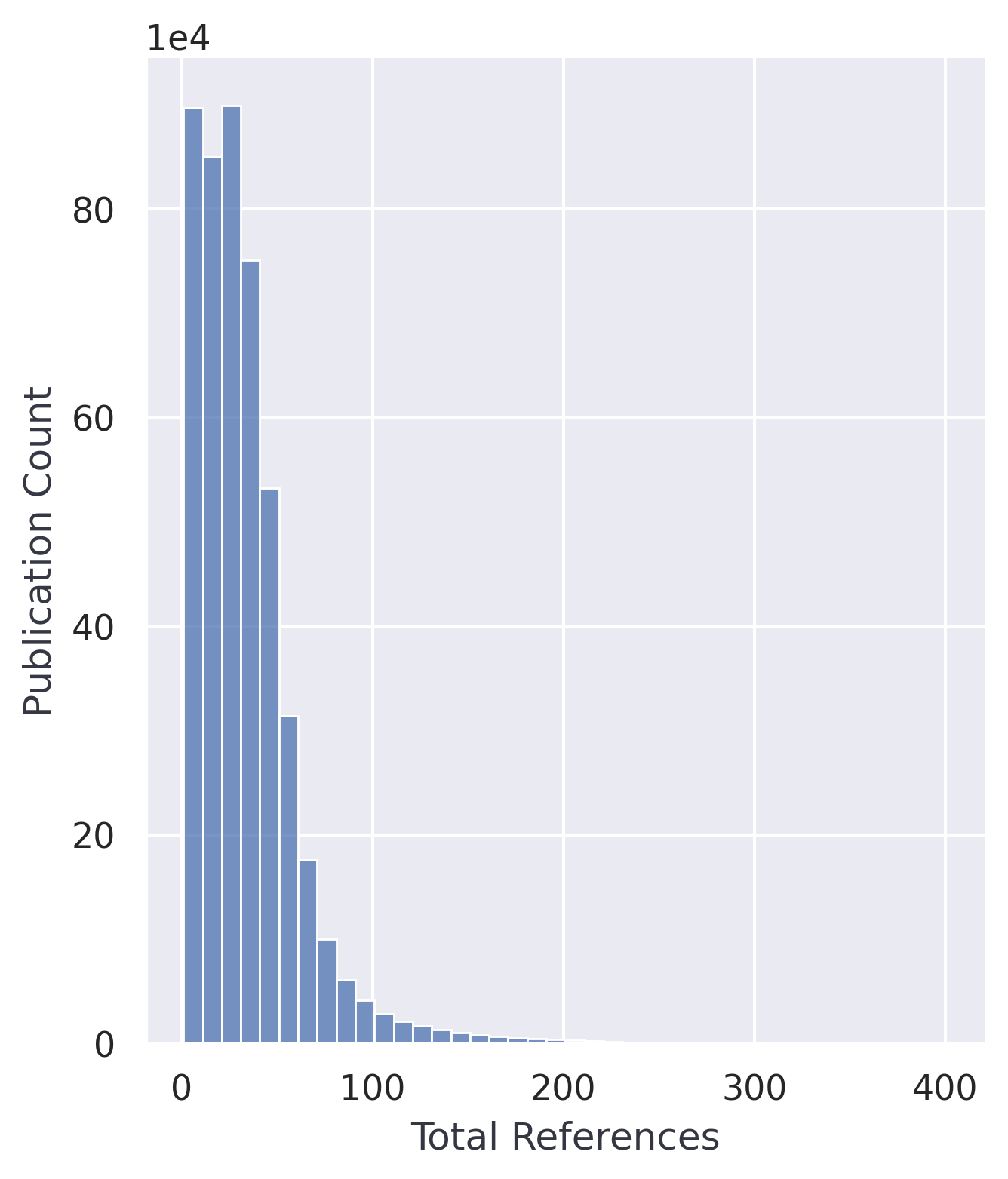}
    \caption{The distribution of publications to total references.There are 1,425 publications with more than 400 references reaching a high of 4,925 references (PMID:9729611). Clipped graph (max total references: 400, bin width: 10)}
    \label{fig:dist_pub_ref}
\end{figure}

\subsection{Data}
For our analysis, we use the articles indexed by PubMed. The 2020 PubMed annual baseline\footnote{\url{https://www.nlm.nih.gov/databases/download/pubmed_medline.html}} consists of more than 30 million publications from which about $5.5$ million also contain their references. Abstracts are available for more than two thirds of the dataset (about 20 million). We narrow down our analysis to articles published since 1990, which limits our dataset to 4,764,686 publications with references. From those, only 368,480 ($7.7\%$) lack an abstract. On average, each publication references $33.2$ other publications indexed by PubMed. Figure \ref{fig:dist_pub_ref} shows the distribution of total references per publication. In total, there are 158,084,383 references, from which 14,973,672 have unique identifiers. While we limit our analysis to specific dates we include referenced publications before 1990. 

\begin{figure}[!b]
    \centering
    \includegraphics[width=0.6\linewidth]{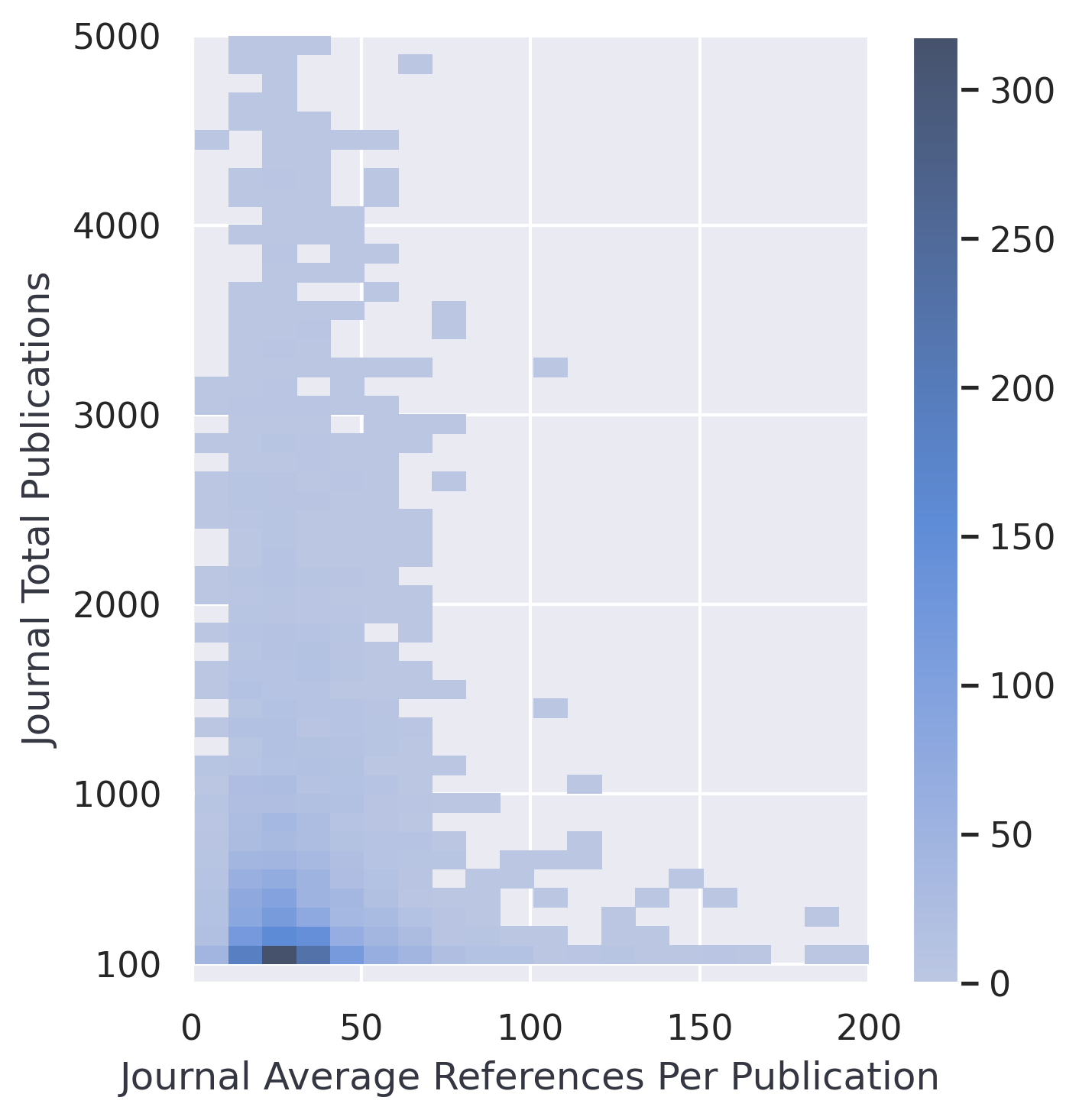}
    \caption{The bivariate distribution of journal total publications and average references per publication. Clipped graph (max total publications: 5000, avg reference per publication: 200, bin width (x,y): 10, 100)}
    \label{fig:bivariate_pub_ref}
\end{figure}

Furthermore, in this dataset, 14,451 unique ISSN identifiers indicate the number of journals in our dataset. The number of publications in these journals ranges from just 1 to more than 200k (i.e PLoS ONE). In this analysis, we consider journals with 100 or more publications, where each publication has at least 10 references. The eligible journals are 3,886 and have 3,810,943 publications. We believe that the journals selected for analysis should have a sufficient number of publications, in order to be able to draw meaningful results. The limit in the number of references was used because some publications only have very few indexed references in PubMed compared to the actual number of references. Figure \ref{fig:bivariate_pub_ref} shows the bivariate distribution of the journal total publications and average references per publication. We notice the majority of journals, regardless of their publications count, reference on average 10-70 other publications while journals with higher reference count have noticeably fewer total publications.

We index our dataset using the Elastic Search Engine. A snapshot of the index containing all the data of the analysis is available upon request. The index includes the full PubMed baseline.

\subsection{Journal Self-Citation}

\begin{figure}[t]
    \centering
    \includegraphics[width=0.7\linewidth]{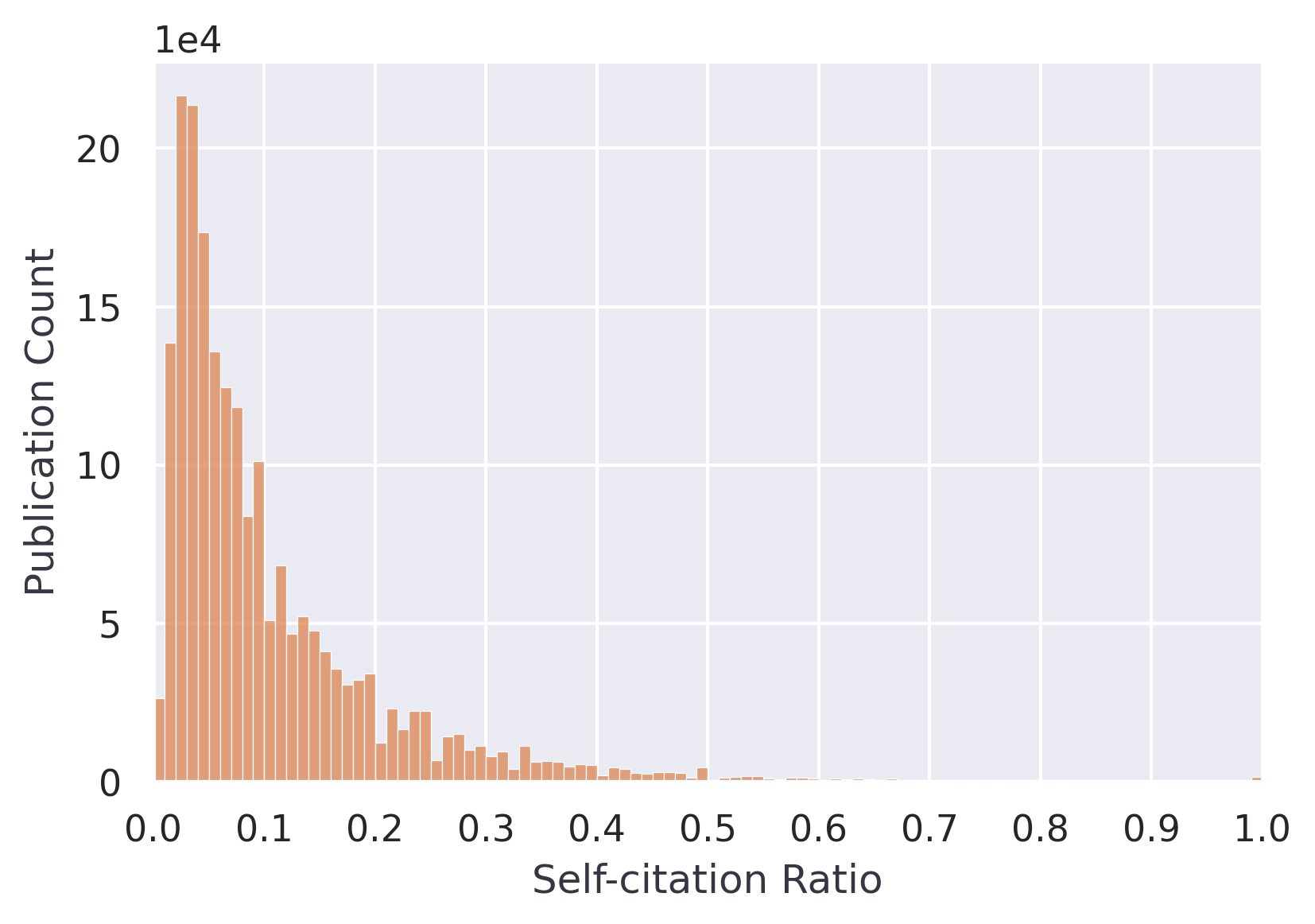}
    \caption{The distribution of the publications and their self-citation ratio. Only publication with more than 10 references and at least one self-citation are included (bin width: 0.01).}
    \label{fig:dist_pub_scratio}
\end{figure}

We consider as a {\em journal self-citation} or {\em journal self-reference} the reference to a publication that has been published by the same journal as the publication that the reference originates from. We compute self-citations for each publication in our dataset. In total, there are 7,588,074 journal self-citations, which correspond to $4,9\%$ of the total references (154,498,212). The number of unique publications with journal self-citations is 2,866,071. Furthermore, these publications have an average of $1.9$ self-citations and a ratio of self-citations to total references of $5.7\%$. 2,079,446 ($53\%$ of the publication have at least one journal self-citation. Figure \ref{fig:dist_pub_scratio} shows the distribution of the publications and their self-citation ratio. 

\begin{figure}[b!]
    \centering
    \includegraphics[width=0.55\linewidth]{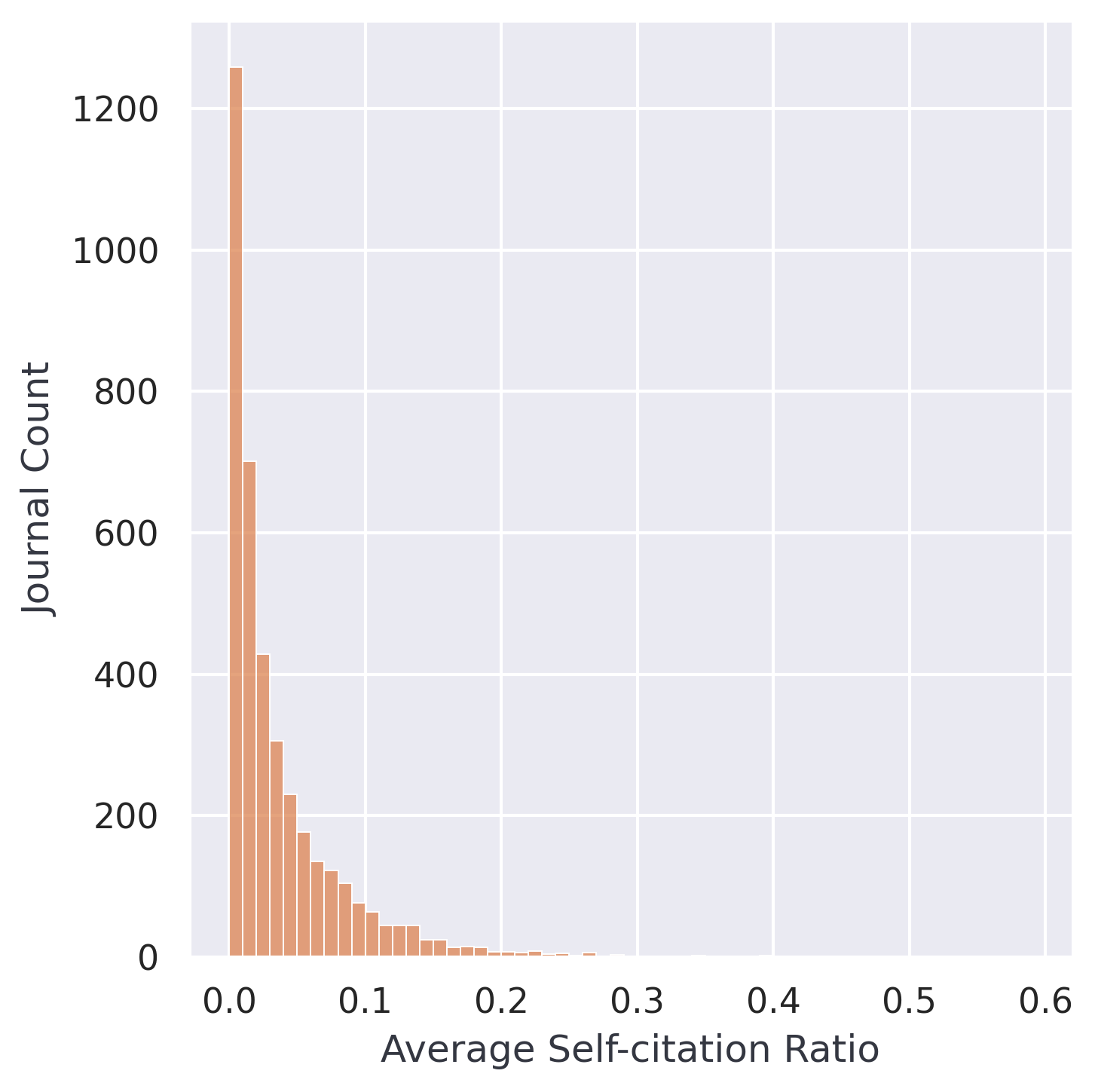}
    \caption{The distribution of the journals and their self-citation ratio. Only publications with more than 10 references are included. Publications with 0 self-citations are also included. (bin width: 0.01).}
    \label{fig:dist_jour_scratio}
\end{figure}

We further compare self-citations across journals. Figure \ref{fig:dist_jour_scratio} shows the journals distribution and their self-citation ratio. The majority of journals ($75.2\%$) have a self-citation rate $<5\%$ while 962 journals have a higher self-citation rate, reaching a maximum of $58.6\%$ (ISSN: 0022-5002). Only 9 journals have a self-citation rate of 0.

\subsection{ReLy Score}

As mentioned earlier, blindly punishing self-citations does not correspond to the reality of citation hacking. Hence, we introduce the ReLy (Relevance Legitimacy) score, which influences the self-citation ratio by using the similarity of publication-reference pairs to measure the relevancy of self-citations to the publication. The score is based upon the idea that publication-reference pair similarities should have approximately the same value, regardless if they are self- or non-self- citations. A low similarity between the publication and a self-citation, when compared to the other references, may indicate an irrelevant reference and question its legitimacy. In this study we use the ReLy score to quantify this influence at journal level, however, it can easily be adapted to other types of self-citations.  

We, first, compute the semantic similarity of a publication-reference pair by extracting semantic vectors of both the publication and the reference. The two vectors are extracted using the BioSentVec \citep{chen2019biosentvec} sentence embedding model on the title and/or abstract of each publication. The BioSentVec model was used due to its state-of-the-art results on biomedical sentence similarity tasks. The cosine similarity of the vectors is then estimated. Thus, the similarity of a publication-reference pair is:

$$
sim(p,r) = cos(\vec{p},\vec{r})
$$

where $p$ ($r$) is the publication (reference), $\vec{p}$ ($\vec{r}$) its vector as extracted by the BioSentVec model,  and $-1<sim(p,r)<1$ since BioSentVec embeddings have both positive and negative values.

For each publication $p$ we compute its average similarity with its references:

\begin{equation}
\label{eqn:similarity}
    sim(p,R) = \frac{\sum_{r \in R} sim(p,r)}{|R|}
\end{equation}

where $R$ is the set of references. Consequently, we can compute the average similarity of self-citations (SCs) and non-self citations (NSCs):

$$sim(p,SC) = \frac{\sum_{r \in SC} sim(p,r)}{|SC|}$$

$$sim(p,NSC) = \frac{\sum_{r \in NSC} sim(p,r)}{|NSC|}$$

where $SC$ and $NSC$ are the sets of self-citations and non-self-citations respectively.

Finally, we define the ReLy score for a publication $p$ as the difference of the NSC and SC similarities multiplied by the self-citation ratio of the publication. The ReLy score is computed as follows:

\begin{equation}
\label{eqn:rely_pub}
    \text{ReLy}_p = \begin{cases}
     \frac{|SC_{p}|}{|NSC_{p}|} \left( sim(p,NSC) - sim(p,SC) \right) & |NSC| > 0 \\
     \hfil 1 & |NSC| = 0
    \end{cases}
\end{equation}

The notion behind this score is that the self-citation ratio of a publication is influenced by the difference of the average similarity of the publication's SCs and the publication's NSCs. In case where all references are self-citations (i.e. $|NSC| = 0$ and $|SC|=|R|$) the score is set to 1. If there are no self-citations the score is set to $0$.  A ReLy score close to 0 indicates that SCs are relevant to the publication and legitimate references. A positive score implies that SCs are irrelevant to the publication and possibly illegitimate references. A negative value, i.e SCs are very similar to the publication, may indicate plagiarism, but this case is left for future investigation.

We can also compute the ReLy score of a journal $J$ by averaging its publications' ReLy score:

\begin{equation}
\label{eqn:rely_jour}
    \text{ReLy}(J) = \frac{\sum_{\forall p \in J}C \times \text{ReLy}_p}{|J|}
\end{equation}

where |J| is the number of publications in a journal and $C$ is a constant used for scaling. In our study, $C=100$ is used.

The ReLy score can easily be adapted to other cases of self-citation since it is only based on what we consider self-citation. Therefore, for author self-citation we only need to define the SC and NSC sets of the corresponding publication. 

Computing the Pearson correlation shows a negative moderate correlation (-0.4049) between the average self-citation ratio and the average ReLy score of a journal.

\section{Results}
\label{sec:results}
This section presents the results of our analysis in the compiled dataset using the semantic similarity between publication and its references and the proposed ReLy score. Since the evaluation of the legitimacy of a reference is a laborious procedure and can't be implemented in large scale we look into different cases by sampling publications with low and high semantic similarity.

\subsection{Semantic Similarity}

\begin{figure}[ht!]
    \centering
    \includegraphics[width=0.55\linewidth]{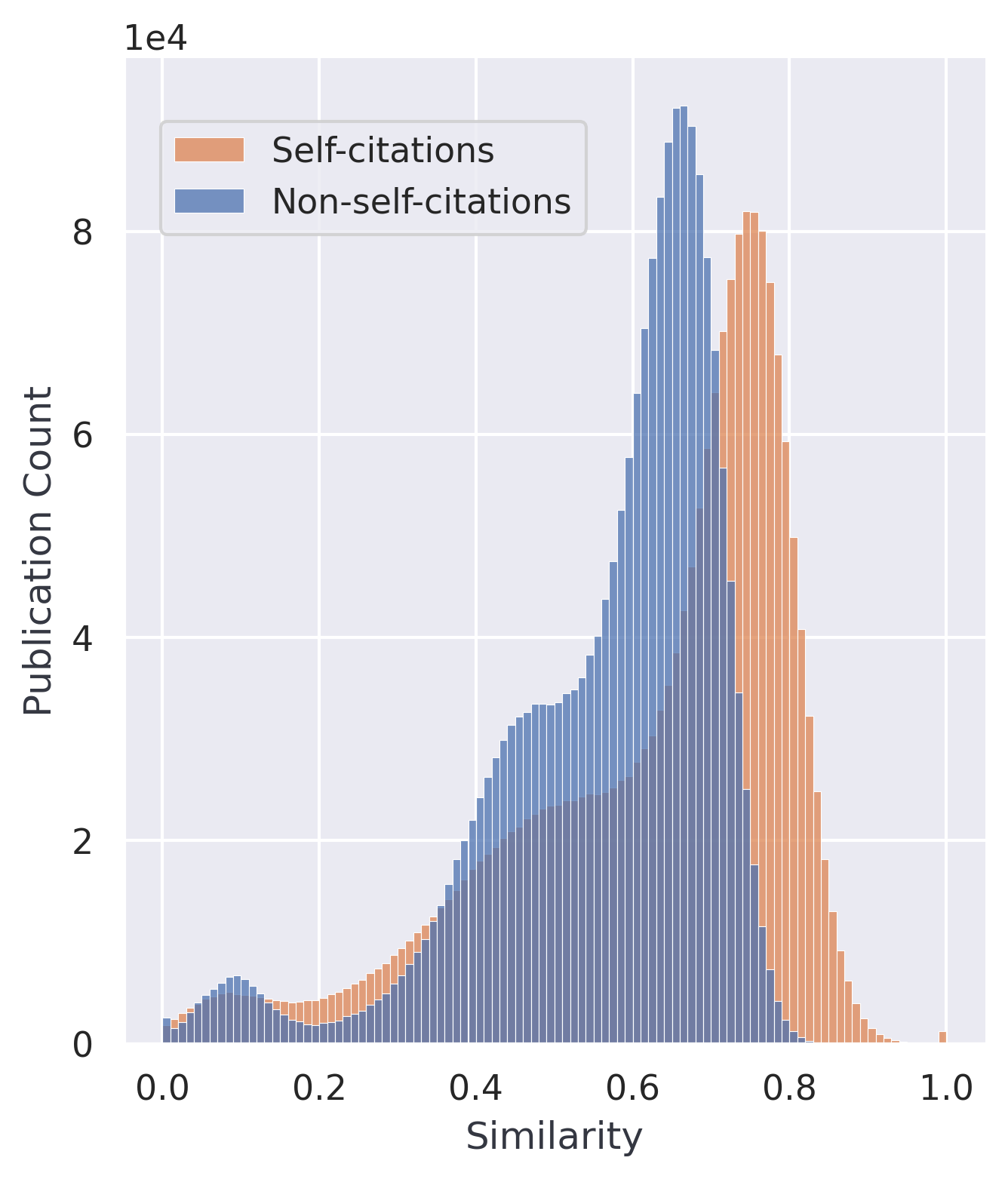}
    \caption{The distribution of the publications and their average semantic similarity with their SCs and NSCs. Publications without SCs are excluded (bin width: 0.01).}
    \label{fig:dist_pub_sim}
\end{figure}

As a first step, we look into the semantic similarity variations between SCs and NSCs of all the publications in our dataset using equation \ref{eqn:similarity}. Figure \ref{fig:dist_pub_sim} shows the distribution of the publications found in our dataset and the average semantic similarity of SCs and NSCs. We notice that SCs have higher similarity to the publication than NSCs. This higher average semantic similarity of SCs is expected since most of the journals concern a specific science field and their publications usually feature similar topics. However, by looking at the semantic difference between NSCs and SCs (i.e. $sim(p,NSC) - sim(p,SC)$)  per publication we find that 513,394  ($24.8\%$) publications have a difference higher than zero and 133,421  ($6.4\%$) publications a difference higher than 0.1. While this is a hint of possible non-relevant references included in a publication, no patterns can be identified.

\begin{figure}[t!]
    \centering
    \includegraphics[width=0.55\linewidth]{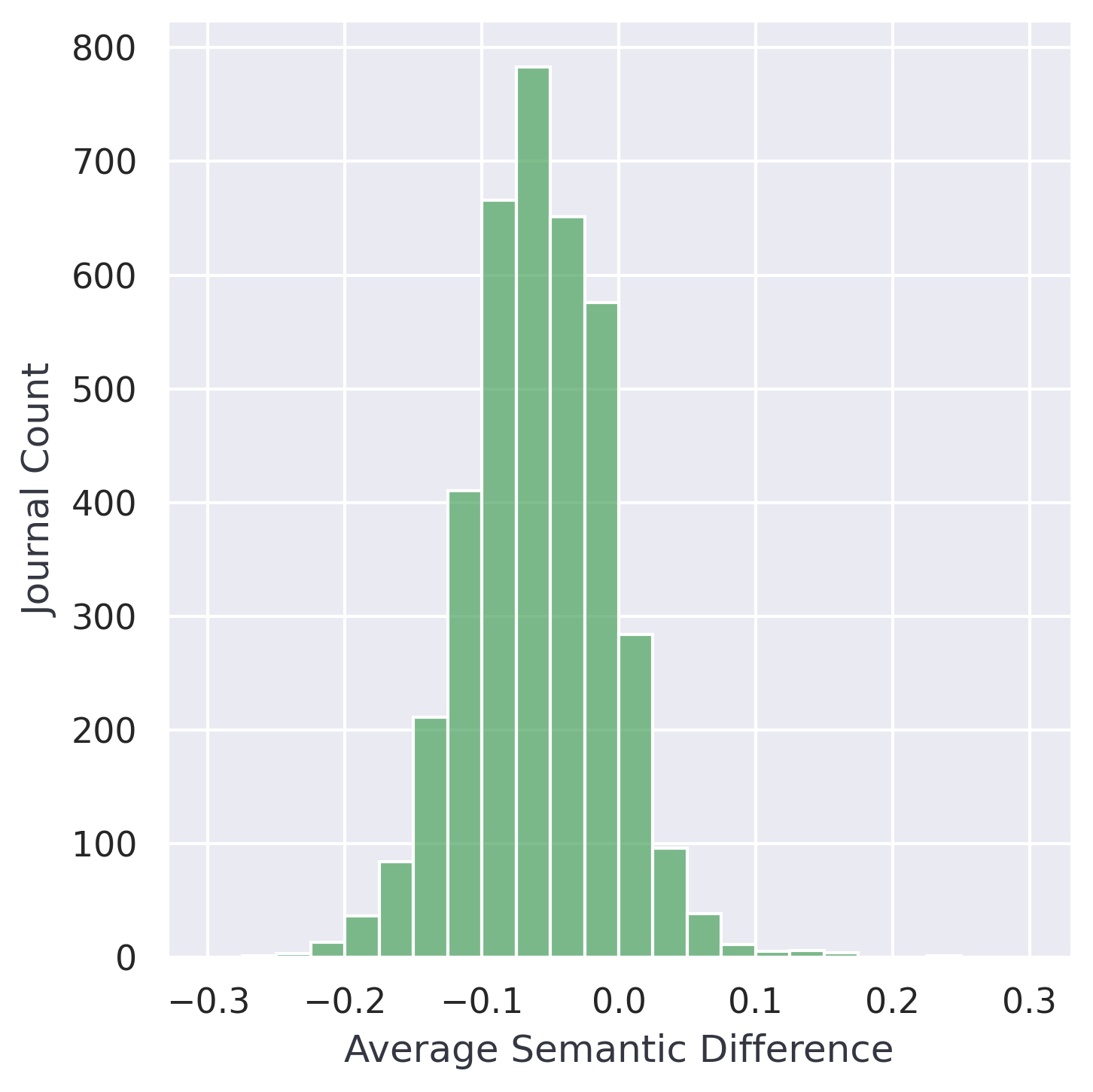}
    \caption{The distribution of the journals and their average semantic difference between NSCs and SCs of their publications (bin width: 0.05).}
    \label{fig:dist_jour_sim}
\end{figure}

Therefore, we further look into the semantic difference between references at journal level. Figure \ref{fig:dist_jour_sim} shows the average semantic difference of journal publications between their NSCs and SCs excluding publications without SCs. Some of the journals in this analysis constantly display semantic differences between NSCs and SCs in their publications. Nevertheless, averaging the similarities of the publications and then again averaging the difference between NSCs and SCs can be misleading. Thus, we look into 5 journals with the highest semantic differences. The journals' ISSNs (SC ratio\footnote{Only on publication with at least one SC.}) are 0091-7451 ($2.3\%$), 2228-6497 ($6.4\%$), 0091-679X ($4.8\%$), 0009-2665 ($2.4\%$), and 2162-1918 ($3.9\%$).

By looking further into journals' articles we noticed that the 0091-7451 and 0091-679X journals have their highest semantic differences on self-citations that are usually review publications or publications with very broad range title (i.e."DNA transformation"). In case of the 2228-6497 journal, all the self-citations with very low similarity were not only journal self-citations but also author self-citations with the same group of others to be part of every publication. Finally, journals 2162-1918  and 0009-2665 while they include SCs to publications with very low similarity, we were unable to determine their legitimacy since our knowledge is out of the context of those journals. Most of their SCs are referring to review articles.

Manually checking for relevant references in publications of a journal is a difficult and time-consuming task. However, by sampling extreme cases of semantic similarity between publications and references we came across some interesting discoveries. We strongly believe that leveraging text representations that hold semantic information can help institutions, authors, and publishers mark potentially illegitimate references of publications and journals. In addition, such models can both used in past and future publications. In past publications, by carrying out analysis on bibliographic sources and in future publications, by verifying the legitimacy of a reference through automatically checking its relevancy to the authoring text. Different thresholds can be drawn by publishers to accommodate such procedures.

\subsection{ReLy Score}

\begin{figure}[ht]
    \centering
    \includegraphics[width=0.57\linewidth]{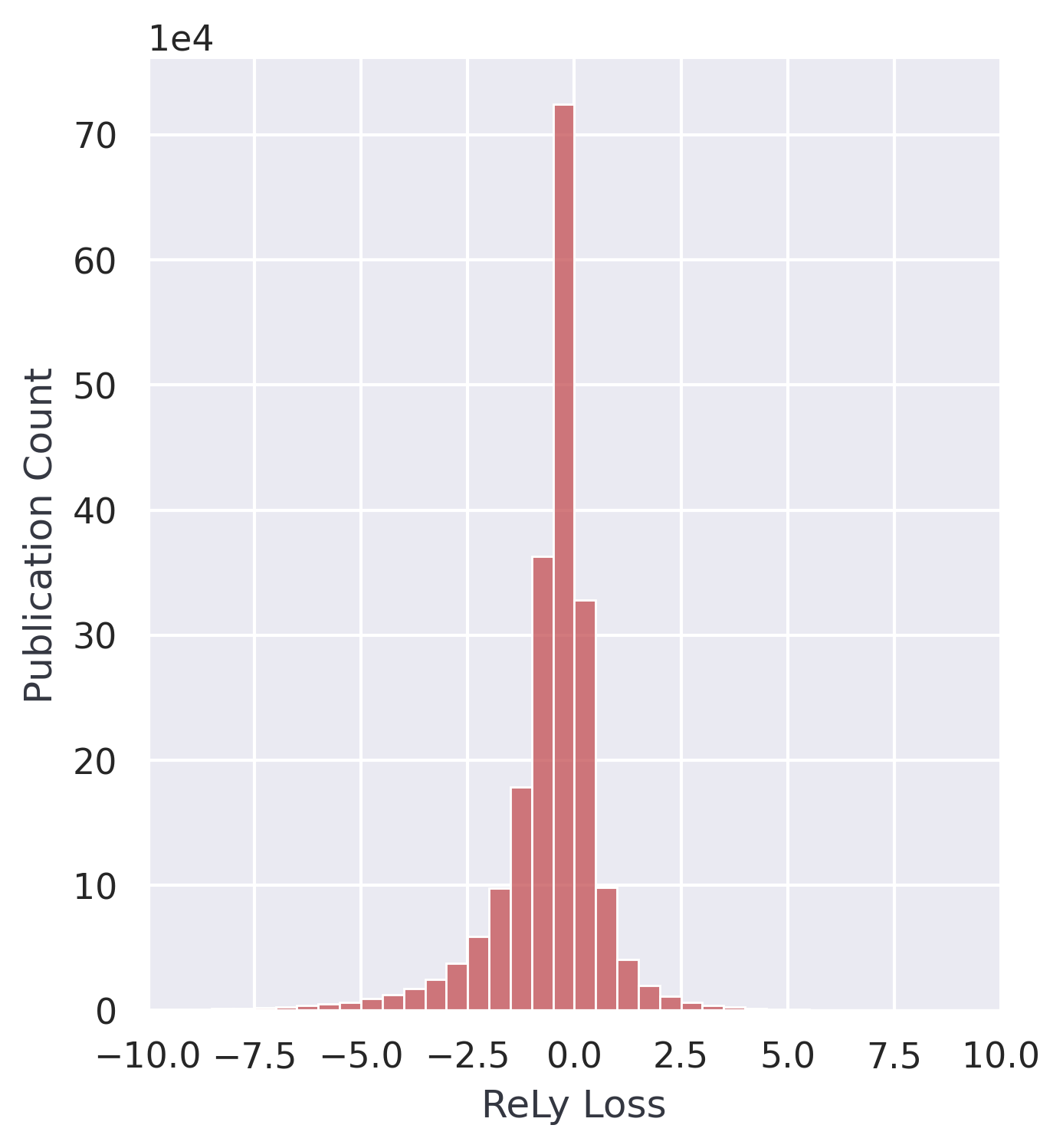}
    \caption{The distribution of the publications and their ReLy score. Clipped graph ($-10.0 < $ReLy $< 10.0$, bin width: 0.5)}
    \label{fig:dist_rely_pub}
\end{figure}

We look further into the semantic similarity of publications and references by computing the ReLy score of all the publications (equation \ref{eqn:rely_pub}) and journals (equation \ref{eqn:rely_jour}) in our dataset. Figure \ref{fig:dist_rely_pub} shows the ReLy score for all publications that have at least one journal SC. The score is multiplied by $C=100$ for comparison reasons. The minimum ReLy score is $\approx -0.7501 (-75.01)$ and the maximum is 1.0 (100.0) where all references are SCs. If we don't include such publications\footnote{Publications with all their references as SCs are excluded since in most cases there were missing references not indexed in PubMed} the maximum score is $\approx 0.4803 (48.03)$. There are 5363 publications with scores less than $-10$ and 2129 publications with scores greater than $10$. Both publication sets show very high SC ratios with 0.6079 and 0.9213 respectively. 

Publications with extreme low ReLy score should suggest that besides the very high self-citation ratio their SCs are very semantically similar to the publication itself, which may pinpoint cases of plagiarism. We manually look at publications with extreme low $(<-10.0)$ ReLy score and sample 5 publications from the 50 publications with the lowest ReLy score. 4 out of 5 publications sampled were from "Acta crystallographica. Section E, Structure reports online" journal which publishes articles on crystal structures and have very similar or almost identical text for different structures. The last publication was from the "Plant physiology" which appears to be a very narrow field journal and its NSCs were to generic publications. Furthermore, we also sampled 10 more publications with lower SC ratios, $<70\%$ and $<50\%$, since the above top-50 had all a ratio above 80\%. The common element between the publications was that their SCs were referring to very similar research, as in the case above. For example, publications on different doses of a drug were tested (Journal of pharmacopuncture), description of genome sequences (Standards in genomic sciences), or series of evaluation publications (Deutsches {\"a}rzteblatt international and Dermatologic Surgery). Additionally, we found a couple of cases where the semantic similarity of a reference was very low due to the absence of abstract text.


On the contrary, publications with extreme high ReLy score should imply that they contain a very high ratio of SCs, but also semantically different references to the publication itself. We manually look at publications with extreme high $(> 10.0)$ ReLy score and sample 5 publications from the top-50 publications with the highest score. 3 out of 5 publications were either characterized as introductory journal articles or editorials. The SCs of one publication were referring to articles of a journal's monograph published in 1943 and, finally, we couldn't determine any specific issues from the last publication sampled due to limited knowledge on the subject (Review on Global health, science and practice journal).  As in the case above, we also sampled 10 more publications with lower SC ratios, $<70\%$ and $<50\%$, to remove any uncertainty about the results of ReLy score with a high SC ratio. Similar cases of publication were discovered, with the most common ones being introductory journal articles, editorials, SCs to specific projects, or SCs to very old publications. Furthermore, we discovered two publications where most of its SCs were to publications including the same author(s), and in one of them, the SCs were also published at the same journal issue.

All in all, the above observations disclose some very interesting cases of possibly illegitimate SCs such as references to editorials and introductory articles. Such cases have also been researched in the past  \citep{heneberg2014parallel,Ioannidis2015,waltman2016review,heneberg2016excessive}. However, by employing the semantic similarity or the ReLy score we can distinguish the legitimacy of such publications more easily and on a large scale without punishing all the publications that belong to those categories. Moreover, we can also uncover cases of malpractice in citations between project publications, very old publications, and citation networks between authors if we further look into those instances using semantic similarity as guidance.

\begin{figure}[!b]
    \centering
    \includegraphics[width=0.57\linewidth]{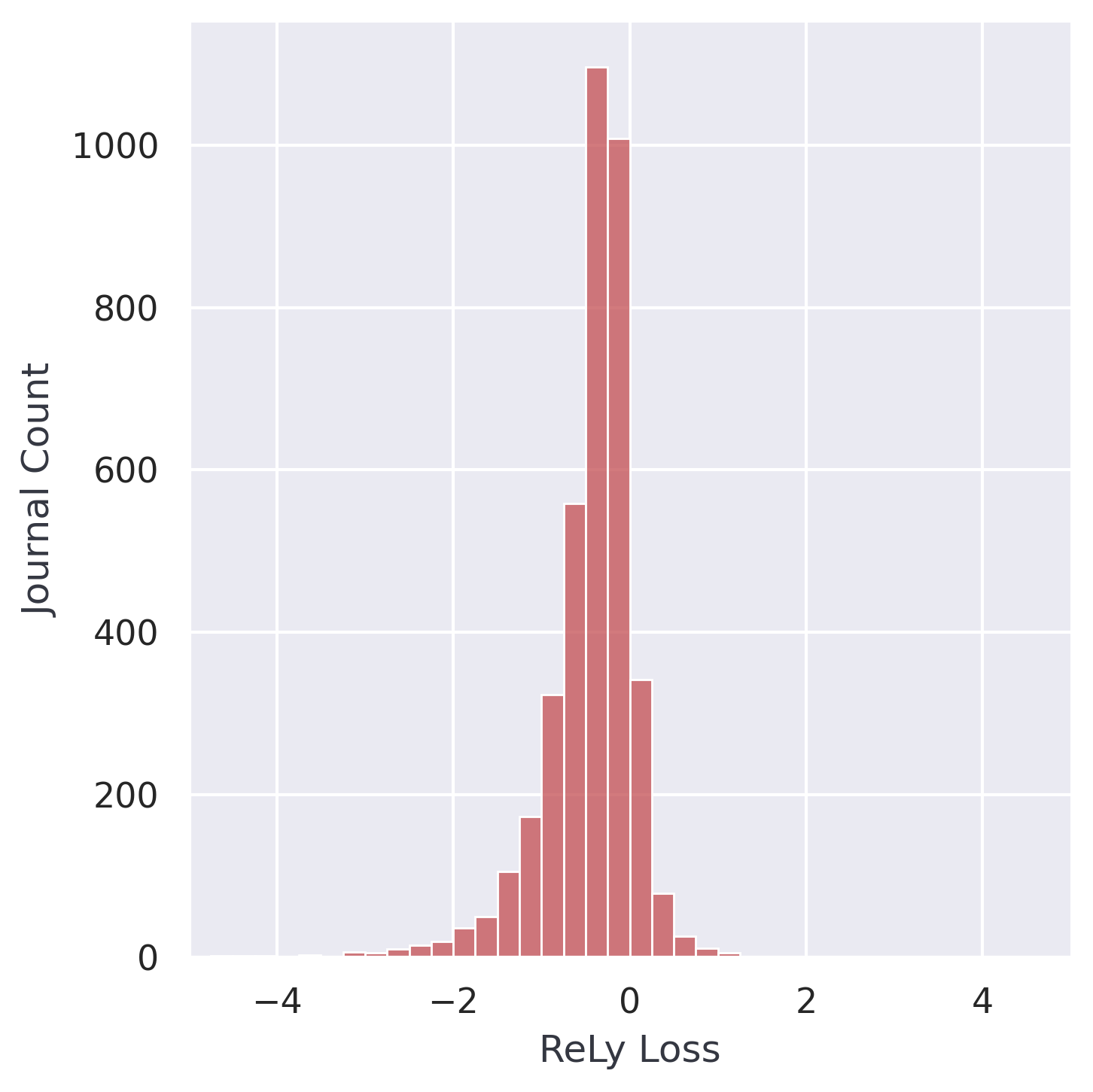}
    \caption{The distribution of the journals and their ReLy score. Clipped graph ($-5.0 < $ReLy $< 5.0$, bin width: 0.25)}
    \label{fig:dist_rely_jour}
\end{figure}

To conclude, we also look into journals using the ReLy score (equation \ref{eqn:rely_jour}). Figure \ref{fig:dist_rely_jour} shows the ReLy score for all journals and their publications that have at least one journal SC, excluding the publications with SC ratio $=1.0$. The average score is $-0.4566$ and there are 41 journals with a ReLy $>0.5$, reaching a high of $2.4406$. The top-5 journals with the highest ReLy are 0100-3984, , 2319-4170, 2255-4971, 1750-0680, 1735-4587. By exploring the publications within the journals we notice that many of the articles included are editorials as also observed before. Also, 3 out of 5 journals are national journals. Finally, many of the publications examined journal SCs were also author SCs.

While ReLy score can determine excessive self-citation malpractice of journals it also allows checking for misconduct cases at publication level. By making publicly available the data of this analysis we believe such cases can be disclosed by researchers of each specialization. We provide the full list of journals with their average semantic similarity and ReLy score in a public repository\footnote{\url{https://github.com/intelligence-csd-auth-gr/rely}} along with all the publications and their computed similarities.

\section{Limitations}
\label{sec:limitations}
The limitations of this study are mainly associated with the data used for our analysis. First, the citation network is limited to the publication indexed by PubMed and to publications whose references are available. Thus, publications may reference articles that are not available in our analysis and the set of a journal's publication may not be complete. Secondly, all the citations included are from the biomedical literature. Thirdly, studies originating in and funded by countries other than the U.S will likely be under-represented in the PMC citation network, as also remarked by Wren et al \citep{Wren2020}. Lastly, some limitations are identified by the use of the sentence embedding model. Specifically, in cases where the abstract is missing from a publication, the similarity score tends to be lower.  

\section{Conclusion \& Future Work}
\label{sec:conclusion}
In this study, we performed a large-scale analysis of journal self-citations in publications indexed by PubMed. We introduce a semantic similarity measure of publication-reference pairs to distinguish legitimate and illegitimate self-citation. The proposed ReLy score does not blindly punish self-citations but influences the self-citation ratio of a publication based on the semantic similarity of its references. We use state-of-the-art sentence embeddings trained on the biomedical literature for text representation. In our analysis, we present the different distributions of the proposed measures and analyze specific cases of possible malpractices. The data used in this analysis are made publicly available for further analysis.

In the future, we plan on extending our research on other types of self-citations such as author self-citations. We also intend to minimize the limitation as much as possible by incorporating more databases and investigating potential drawbacks of the sentence embedding model used.

\section*{Acknowledgment}
This research  is co-financed by Greece and the European Union (European Social Fund- ESF)  through the Operational Programme Human Resources Development, Education and  Lifelong Learning in the context of the project Strengthening Human Resources Research Potential via Doctorate Research (MIS-5000432), implemented by the State Scholarships Foundation (IKY).

\bibliographystyle{unsrtnat}
\bibliography{references}  






\end{document}